# Resonant structure in the disks of spiral galaxies, using phase-reversals in streaming motions from 2D Hα Fabry-Perot spectroscopy

Joan Font (1,2), John E. Beckman (1,2,3), Benoît Epinat (4,5) Kambiz Fathi (1,6,7), Leonel Gutiérrez (1,8), Olivier Hernandez (9).

1. Instituto de Astrofísica de Canarias, c/Vía Láctea, s/n, E38205, La Laguna, Tenerife, Spain: jfont@iac.es, jeb@iac.es.
2. Departamento de Astrofísica. Universidad de La Laguna, Tenerife, Spain.
3. Consejo Superior de Investigaciones Científicas, Spain.
4. Laboratoire d'Astrophysique de Toulouse-Tarbes. Université de Toulouse, CNRS, 14 Avenue Édouard Berlin, F-31400, Toulouse, France: benoit.epinat@ast.obs-mip.fr.
5. CNRS, IRAP, 9 Av. Colonel Roche, BP 44346, 31028, Toulouse, Cedex 4, France.
6. Department of Astronomy, University of Stockholm, AlbaNova, 10691, Stockholm, Sweden: kambiz@astro.su.se.
7. Oscar Klein Centre for Cosmoparticle Physics, Stockholm University, 10691 Stockholm, Sweden.
8. Instituto de Astronomía, Universidad Nacional Autónoma de México, Apdo. Postal 877, 22800, Ensenada B.C., México: leonel@astrosen.unam.mx.
9. Département de Physique et Astronomie, Université de Montreal, CP. 6128, Succ. Centre ville, Montréal, QC H3C 3J7, Canada: hernandez@astro.umontreal.ca.

ABSTRACT

In this article we introduce a technique for finding resonance radii in a disk galaxy. We use a two-dimensional velocity field in Hα emission obtained with Fabry-Perot interferometry, derive the classical rotation curve, and subtract it off, leaving a residual velocity map. As the streaming motions should reverse sign at corotation, we detect these reversals, and plot them in a histogram against galactocentric radius, excluding points where the amplitude of the reversal is smaller than the measurement uncertainty. The histograms show well-defined peaks which we assume to occur at resonance radii, identifying corotations as the most prominent peaks corresponding to the relevant morphological features of the galaxy (notably bars and spiral arm systems). We compare our results with published measurements on the same galaxies using other methods and different types of data.

I. INTRODUCTION.

The density wave model (Lindblad 1961, Lin & Shu 1964) is the scenario of reference for understanding the prevalence of spiral structure in galaxies. A key parameter is the corotation radius, where the angular velocities of the spiral density pattern and of the matter in the disk are equal. As our knowledge of disk galaxies has grown, observers are finding that there may be more than one pattern speed, for example one associated with a bar, and another with the arm structure further out in the disk (see, e. g. Meidt et al. 2009). The importance of the measuring the bar pattern speed has been enhanced by predictions that a slow bar would be consistent with the braking by a live dark halo, while a fast bar would not (for a recent treatment see Villa Vargas et al. 2009, and references therein).

Methods to measure pattern speeds can use morphology, kinematics, or both. Perhaps the best known, due to Tremaine & Weinberg (1984) combines stellar column density, via photometric imaging, and line of sight velocity, via long slit spectra parallel to the disk major axis. It has been used on the stellar component of 17 galaxies (see Corsini 2011 for a review). As it assumes continuity of the emitting source, opinion is divided on whether it can also be applied using interstellar gas as shocks, phase conversions of gas and star formation may each weaken this assumption, but Zimmer et al. (2004), applied it to CO emission line maps of 3 galaxies and Rand & Wallin (2004), applied it to 6 more, while Hernandez et al. (2005), used an Hα emission line map of NGC 4321 and Fathi et al. (2009), used similar maps for 10 galaxies. Other methods used have included predicting the HI velocity field using measured gravitational potential distribution (Sanders & Tubbs, 1980; England et al. 1990; Garcia-Burillo, et al. 1993), to derive the pattern speed, and the method of Canzian (1993) which places corotation where the azimuthal non-circular motion pattern changes from singlet to triplet. This was applied to an HI velocity field by Canzian & Allen (1997) to NGC 4321, and we will compare their work to ours below. Sempere et al. (1995) presented a variant on the Canzian method, also related to the technique described here. Buta & Zhang (2009) used the potential-density phase shift to find corotation for 153 galaxies, which we will use for comparison with our results. In addition Vega Beltran et al. (1998) identified resonances using rings, Athanassoula (1992) showed that dust lanes marked offset shocks in gas flows, and Puerari & Dottori (1997) pinned down corotation using two-color photometry. For an overview of methods for bar pattern speeds see Rautiainen et al. (2008).

In the present article we present a method for finding corotation using the change in sense of the non-circular velocity components, relying on the high information density of kinematic maps in H$\alpha$ and HI. The method detects such changes at more than one radius, which for the present we identify as multiple corotations, corresponding to multiple pattern speeds. We give the results for a set of galaxies chosen because their corotation radii were reported previously, allowing us to make comparisons.

II. THE TECHNIQUE DEVELOPED TO EXPLORE THE RESONANCES

The method we have adopted to explore the resonant structure of a galaxy is to find the galactocentric radii at which the gas response shows a 180º change of phase. This is a more simple-minded way of probing for resonances than most of the previously used techniques, but as we will see it is powerful. It has become possible to use this because the observational material we have at our disposal has much finer angular resolution than is customary for observations made using the interstellar gas component, for the obvious reason that we are using optical wavelengths, whilst until now the HI or CO maps used have had typically coarser resolution by an order of magnitude (this situation will change radically when ALMA is in regular operation). We use data cubes in H$\alpha$ obtained using Fabry-Perot interferometry, which yield maps of galaxies in surface brightness, radial velocity, and velocity dispersion using the emission from their ionized interstellar gas. The technique is applied by first deriving the rotation curve of the galaxy: the circular velocity as a function of galactocentric radius, as the circular velocity is traced by the rotation velocity of the ionized gas which is assumed to follow the circular orbits. We then subtract off the rotational velocity component, suitably projected, from the complete map of radial velocities, yielding a map of "residual velocity", the radially projected component of the departures from systematic galactic rotation, across the face of the galaxy. The next step is to identify all the points where the residual velocity is zero, and plot them on the face of the galaxy. In an ideal, noise-free system these zeros would correspond to the places where the streaming motions induced by a density wave system are zero because of a change in phase of 180º corresponding to a change in sign of the flow vector of the streaming motions as pointed out by Kalnajs (1978). These changes should occur at resonance radii. In practice it is clear that in any non-ideal system these zeros will occur at many places where noise causes a measured transit from negative to positive velocities, or vice-versa. In order to optimize the reliability of the measurement we then excluded those zeros from the map where the detected change in velocity across the zero, either positive or negative, was smaller in magnitude than the uncertainty in velocity. The following step was to plot the number of zeros in radial annuli (effectively elliptical because of the inclination of the galaxy disk), against radius. These graphs showed peaks at specific radii, often very marked and with radii easy to determine. The question of which peak corresponds to which resonance is not trivial to resolve. In the present article we use two methods: inspection of the morphology of the galaxy, followed by fitting the peaks to radial plots of $\Omega$, $\Omega\pm\kappa/2$, $\Omega\pm\kappa/4$, having selected which peak should correspond to corotation, and comparison of our results with previous results for the same galaxies obtained previously using other methods.

III. THE OBSERVATIONAL DATA.

The data used here come from three separate sources: (a) the GHASP (Gassendi H Alpha Survey of Spirals; Epinat et al. 2008) survey of star forming galaxies using a scanning Fabry-Perot interferometer on the Cassegrain focus of the 1.93-m telescope at the Observatoire de Haute Provence (data, which include observational parameters, rotation curve and residual velocity map, are available at the Fabry-Perot database: http://fabryperot.oamp.fr/FabryPerot/index.jsp). The detector consists of a CCD camera together with an image photon counting system (IPCS) which leads to zero read out noise (Gach et al. 2002). Depending of the observation, the field-of-view of the instrument is 5.8 arcmin$^2$ or 4.1 arcmin$^2$ allowing to obtain images with a pixel size of 0.68 arcsec or 0.96 arcsec, respectively. Seeing limited angular resolution is ~3 arcsec (FWHM) and spectral resolution is ~30 km/s. NGC 4321 does not belong to this survey although it was observed at the same telescope with the same instrument (Chemin et al. 2006). (b) the commissioning run of GH$\alpha$FaS (Galaxy H$\alpha$ Fabry-Perot System; Hernandez et al. 2008) in July of 2007, a new generation scanning Fabry-Perot interferometer-spectrometer on the 4.2m William Herschel Telescope, Observatorio del Roque de los Muchachos, La Palma, Canary Islands. The instrument is used at the Nasmyth focus of the telescope, for practical ease of operation, leaving a field-of-view no larger than 3.5 arcmin$^2$ which, with the CCD camera with IPCS used, gives a pixel size of 0.398 arcsec. The ~1.5 arcsec of seeing (FWHM) achieved in that run limits the angular resolution, and the spectral resolution in velocity was ~16 km/s. (c) the VIVA (VLA Imaging of Virgo in Atomic Gas survey (where VLA is the Very Large Array)); Chung et al. 2009) consisting of HI data of 53 late type galaxies with a typical spatial resolution of 15 arcsec (FWHM) and velocity resolution of ~10 km/s; data are available at the VIVA webpage: http://www.astro.yale.edu/viva/.

IV. RESULTS

The results presented here are for a sample of 8 galaxies (for one of which, NGC 4321, there are both H$\alpha$ and HI velocity maps, while for the other 7 only H$\alpha$), chosen because their resonance radii had been published previously; the references are specified in the footnote to Table 1. In Figure 1 we have plotted the number of zero velocity crossings in annuli of equal width as a function of galactocentric radius, normalized to unity at the highest peak in each distribution. In all cases there are well defined peaks, at radii which we can take as indicating corotation. The peaks are narrow enough for us to make clean measurements of these radii, as shown in the figure, and we present the resulting sets of radii in Table 1. These are presented sequentially in order of increasing radius, in the left hand section of the table with their errors, which take into account the standard deviation of Gaussian fit to each peak and angular resolution. In the right hand section we include, for comparison, measurements described by their authors as corotation radii. The principal corotation radii measured using our method are noted in bold face. In the majority of the galaxies these were the single most prominent maxima in our plots (NGC 428, NGC 3344, NGC 3726, NGC 5921) or there were several comparably strong peaks (NGC 7479, NGC 7741) which are all in bold face. The published values used for our comparisons were obtained by several of the methods described in the Introduction. We will discuss them, galaxy by galaxy.

(a) Corotation using H$\alpha$ velocity fields.

NGC 428. We find 5 resonance radii for this object (Figure 1a). The resonance at 59 arcsec gives by far the strongest signal, and we assign this as corotation; it is located at a 1.2 times the radius of weakly defined bar. The only literature value, due to Buta & Zhang (2009), is 54.8 arcsec, which may be compatible within the error bars.

NGC 3344. Of our 4 resonance peaks (Figure 1b) that at 102.8 arcsec is by far the strongest, so we assign corotation to this radius. It is clearly not associated with the central weak bar, so we assume it is a feature of the spiral arm system. Meidt et al. (2009), using CO velocities, derived a pattern speed via the Tremaine-Weinberg method, which we have used to determine the radii of corotation, and of the two Lindblad and the two 4:1 resonances. The agreement for corotation (102.4 arcsec) is excellent, and the radii of the 4:1 resonances, 57.7 arcsec and 126.0 arcsec, agree very well with two of our peaks (58.7 and 127.3 arcsec).

NGC 3726. In Figure 1c we show 5 resonance peaks, of which the peak at 100.1 arcsec is clearly the strongest, so we assign corotation to this value. Inspection shows that this is around 1.25 times the length of the main bar. The innermost peak appears to be associated with an inner boxy mass concentration of the bar, while the two outermost peaks are disk features. Only one of Buta & Zang's three corotations at 65.2 arcsec is compatible with our results.

NGC 5427. As seen in Figure 1d the two outermost peaks are strong, and we have selected the stronger, 47.3 arcsec for corotation, with no clear help from the morphology. The peak at 23.5 arcsec appears associated with the inner bar. Of Buta & Zang's 3 corotation values, the third at 63.3 arcsec coincides with our fourth peak.

NGC 5676. In figure 1e we see 4 resonance peaks, and have assigned corotation to the outermost, most intense, peak at 76.5 arcsec. The inclination makes it hard to find correspondence with a specific feature. The peak at 18.4 arcsec appears associated with the small bar. None of Buta & Zang's corotation radii coincide with ours.

NGC 7479. As seen in Figure 1f we show 6 peaks for this very strongly barred starburst galaxy (Laine & Gottesman, 1998). Using criteria of peak intensity, and in agreement with other authors (see table 1), we selected 55.8 arcsec and 98.3 arcsec as two corotation radii, one associated with the bar, and the other with the disk. The innermost peak, at 13.2 arcsec, belongs to the central mass concentration.

NGC 7741. Figure 1g shows 4 peaks, of which 3 are comparably strong, and we assign to each of these a corotation. The peak at 36.8 arcsec is linked with the central mass concentration of the bar, that at 66.3 arcsec is associated with the bar, (at ~1.3 the bar radius), while that at 95.3 arcsec is due to the disk system. Buta & Zang's velue of 51.9 arcsec would be reproduced by a smoothed/averaged contribution of our inner two peaks, while Fathi et al. (2009) find corotation consistent, within the error bars, with our outermost peak.

(b) **Corotation in NGC 4321** using HI and Hα: comparison with Canzian´s method

In Figure 2a we show the 5 peaks tabulated in Table 2. Of these the innermost peak at 31.8 arcsec is associated with the central mass concentration (a small bar), the second peak at 97.0 arcsec is linked to the weak long bar. We have termed both of these corotations. Both are also detected directly in HI (Figure 2b), and Canzian & Allen (1997) range of 88-108 arcsec includes the 97.0 peak. In Figures 2c and 2d we show, respectively, the residual velocity maps in Hα and HI (used to prepare Figures 2a and 2b), which both show the transition from a single azimuthal cycle of velocity transitions in the central zone to a triple cycle in the outer zone. The radius of the peak close to 100 arcsec in the histograms of Figures 2a and 2b, marked on the maps (Figures 2c and 2d), is that of the singlet-triplet transition. Our technique thus picks out this transition, but more sharply, using the Hα field. It is notable, however, that Hernandez et al. (2005), using the Tremaine-Weinberg method on the same Hα velocity map that we have used here, find two of our five corotation radii (97 and 150 arcsec), with exceptionally good agreement, within the limits of the rather small error bars in both sets of results.

Conclusions.

We have developed a method, using maps at high angular resolution, of the radial velocity fields of disk galaxies, which allows us to determine their resonant structure. The principle is direct, finding the maxima in the radial distribution of resolution elements harboring a transition through zero in the residual velocity map, after subtracting off the rotation curve. Transitions where either the positive or negative velocity at either side of the zero is less than the measurement uncertainty were excluded, reducing noise and eliminating projection effects near the minor axes. The peaks are generally well-defined in radii, and narrow. Based on the general theoretical prediction that at a resonance we should find this kind of phase changes, we selected the strongest peaks as candidates for corotation. The technique is simple, not needing combinations of parameters integrated over the disk. Comparison of our results with values found by means of other methods gives quite satisfactory agreement in some cases only. Our values are usually more sharply defined, in some cases due to the nature of the other techniques, in others because the authors used maps with lower spatial resolution. A guide value for the uncertainty in defining corotation using our method, a mean of our measurements with H$\alpha$, is 5% rms. This compares favorably with 13% derived by Meidt et al. (2009) for NGC 3344, using CO and HI velocity maps with an angular resolution of 7 arcsec, with 12% for NGC 4321 derived by Hernandez et al. (2005), using the Tremaine-Weinberg method on an H$\alpha$ velocity map, and with 12%, which we obtain using our method but with lower angular resulution HI data from VIVA (Chung et al. 2008). We believe that the method presented here is a significant step forward in quantifying the resonant structure of disk galaxies, but understand that there are issues to resolve for this method and for the others, related to the nature of the resonances, their structure, and their lifetimes.

Acknowledgements

This work was supported by projects AYA2007-67625-CO2-01 of the Spanish Ministry of Science and Innovation, and project 3E 310386 of the Instituto de Astrofísica de Canarias. We thank Sharon Meidt, Scott Tremaine, and Phil James, for very useful discussions, and the anonymous referee for valuable comments.

FIGURES & TABLES

Table 1
Comparison of resonance radii for sample galaxies with Hα velocity maps

| NGC | Type | | I | II | III | IV | V | VI |
|---|---|---|---|---|---|---|---|---|
| | | | | | Resonance radii (arcsec) | | | |
| 428 | SAB(s)m | This work | 5.8±2.2 | 17.6±3.1 | **59.5±3.9** | 115.6±2.2 | 144.7±2.2 | |
| | | Other works | | | 54.8[a] | | | |
| 3344 | SAB(r)bc | This work | | 58.7±3.8 | 74.4±1.9 | **102.8±2.3** | 127.3±2.3 | |
| | | Other works | 26.2±5.2[b] | 57.7±10.4[b] | | 102.4±13.1[b] | $126.0^{+15.7}_{-10.4}$ [b] | $160.1^{+18.4}_{-13.1}$ [b] |
| 3726 | SAB(r)c | This work | 25.2±2.8 | 63.0±2.9 | **100.1±3.2** | | 122.4±3.2 | |
| | | Other works | 21.9±1.0[a] | 65.2±2.1[a] | | 114.1[a] | | |
| 5427 | SA(s)c | This work | | 15.7±2.3 | 23.5±0.8 | | **47.3±2.1** | 64.5±0.9 |
| | | Other works | 7.3±0.2[a] | | | 38.0[a] | | 63.3[a] |
| 5676 | SA(rs)bc | This work | 18.4±1.9 | 32.1±1.7 | **54.4±2.4** | | 76.5±4.6 | |
| | | Other works | 23.2[a] | 38.2[a] | | 61.2[a] | | |
| 7479 | SB(s)c | This work | 13.2±2.8 | 35.3±5.2 | **55.8±4.8** | 80.8±4.9 | **98.3±3.1** | 123.1±3.0 |
| | | Other works | 6.3[a] | 23[f] | 57.7[a];50[d];55[f];57[g] | 85[e] | $94^{+6}_{-25}$ [c];95[d];74-106[h] | |
| 7741 | SB(s)cd | This work | **36.8±3.6** | | **66.3±1.9** | 77.4±4.5 | **95.3±2.7** | |
| | | Other works | | 51.9±6.8[a] | | | $109^{+?}_{-17}$ [c] | |

Notes. For each object, upper row: resonance radii (in arcsec) measured using the technique introduced in the present article. The columns are sequential from left to right, in order of increasing galactocentric radius, enumerated separately by order. Suggested corotation radii are in boldface. Lower row: measurements of the corotation radius obtained by the authors identified in the footnote, using a variety of methods.
[a] from potential-density phase-shift method (Buta & Zhang 2009)
[b] calculated from pattern speed of Meidt et al. (2010) applying our rotation curve.
[c] from Tremaine-Weinberg method on Hα data (Fathi et al. 2009). (These authors could not obtain an upper uncertainty limit for the corotation of NGC 7741 with the data available to them).
[d] from simulations (Laine et al. 1998)
[e] from band crossing method (del Rio & Cepa 1998)
[f] from morphological method (Puerari & Dottori 1997)
[g] from simulations (Sempere et al. 1995)
[h] from morphological inspection (Elmegreen & Elmegreen 1995).

Table 2
Comparison of resonance radii of NGC 4321

| I | II | III | IV | V | VI | VII | Reference |
|---|---|---|---|---|---|---|---|
| 33.1±10.6 | 76.5±6.8 | 108.8±10.3 | | | 166.3±6.0 | | HI[a] |
| **31.8±3.4** | | **97.0±2.2** | 123.5±2.1 | 147.9±10.7 | | 188.2±9.5 | Hα[b] |
| | | 88-108 | | | | | 1 |
| | | 110 | | | | | 2 |
| | 71.8 | | 123.1 | | | | 3 |
| 42 | | 97 | | 150 | | | 4[c] |

Notes. [a,b] Resonance radii (in arcsec) measured using the technique introduced in the present work using velocity map in HI by Chung et al. 2009 and in Hα by Chemin et al. 2006. [c] Calculated from results of Hernandez et al. 2004 who used the Tremaine-Weinberg method on Hα observations. The three radii correspond to three corotations: nuclear, bar and spiral, respectively.
**References**. (1) Canzian & Allen 1997; (2) Sempere et al. 1995; (3) Elmegreen et al. 1989; (4) Hernandez et al. 2004.

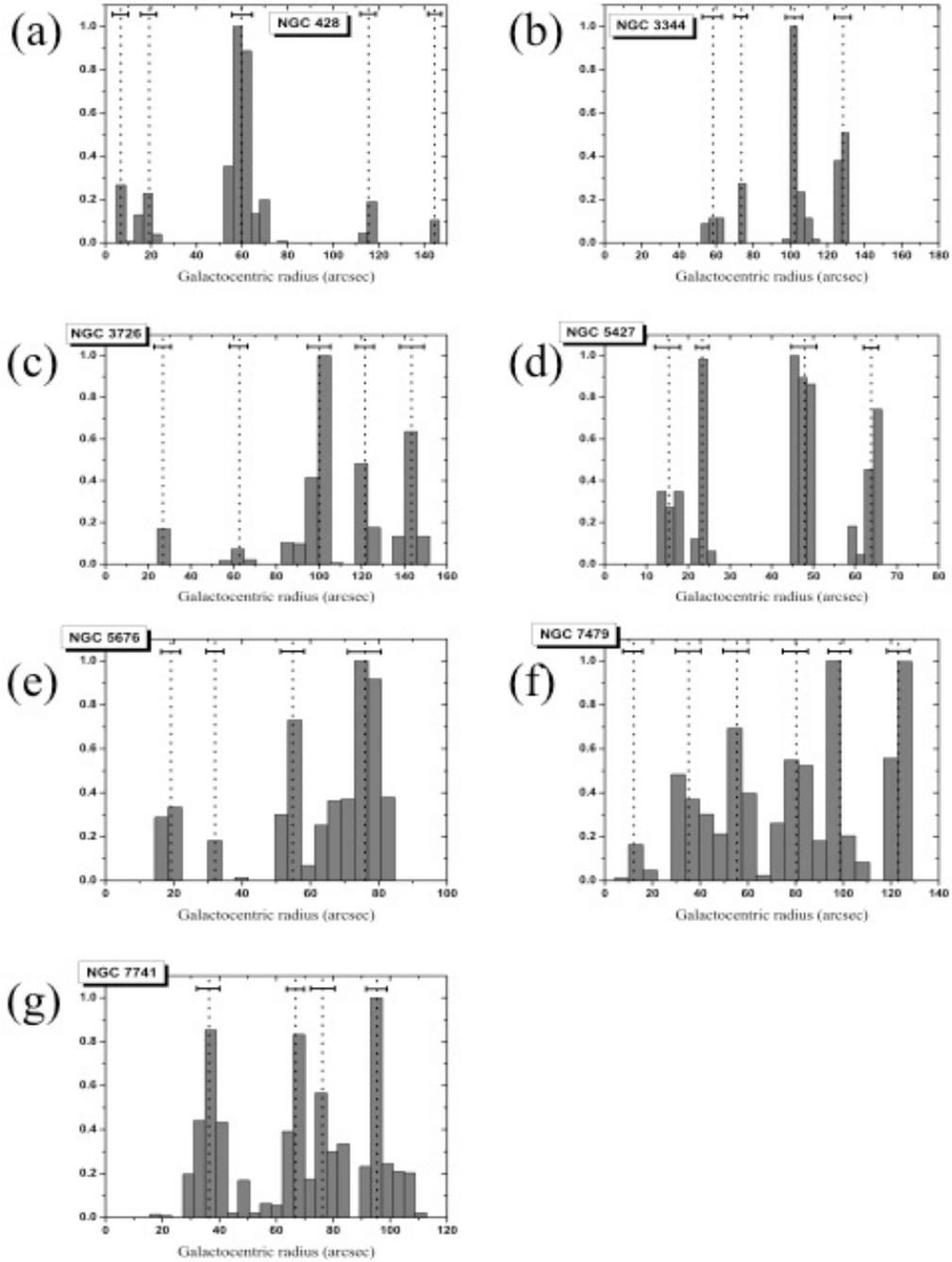

**Figure 1.** Normalized histograms of the number of zero crossings in the residual velocity fields of the galaxies measured, as functions of galactocentric radii (ie. radii in the galactic plane). In all cases the graphs show well defined maxima, which we associate with the effects of resonances corresponding to density wave structure, generally denoting corotation radii. (Vertical dotted lines locate the resonance radii and horizontal segments on top of the histogram show the uncertainty bar associated to each radius).

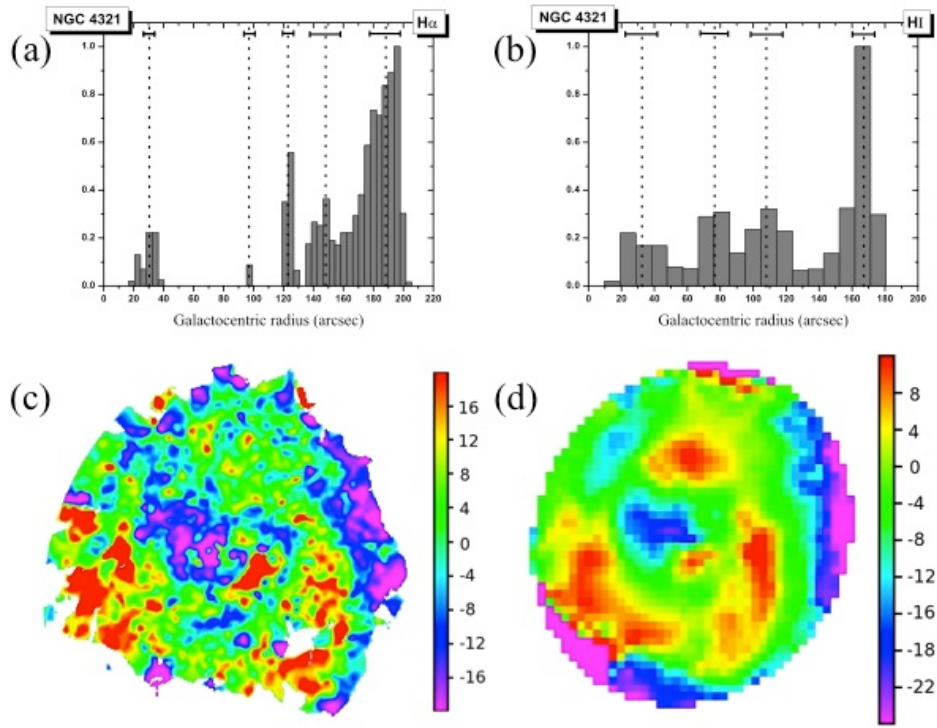

**Figure 2.** Panels a & b. Normalized histograms of the density of sign changes in the residual velocity maps of NGC 4321 (M100) as a function of the radius in the galactic plane (galactocentric radius). To the left, from the Hα velocity field (Chemin et al. 2006), to the right, from the HI velocity field (Chung et al. 2009). Resonance radii and its uncertainty bar are shown as vertical dotted lines and horizontal segments, respectively. Panels c & d. Residual velocity maps in Hα (left) and HI (right), showing the transition from singlet to triplet structure predicted by Canzian (1993). (Velocity key to the right in km s$^{-1}$). The peaks close to 100 arcsec in the histograms occur at this transition.